\documentclass{ws-ijmpa}
\def\M{M_{K\overline K}}   

\begin{document}

%\markboth{Authors' Names}
%{Instructions for Typing Manuscripts (Paper's Title)}

\catchline{}{}{}{}{}
%
%%%%%%%%%%%%%%%%%%%%%%%%%%%%%%%%%%%%%%%%%%%%%%%%%%%%%%%%%%%%%%%%%%%%

\title{Analysis of high energy $K^+K^-$ photoproduction on hydrogen}

\author{\footnotesize \L . BIBRZYCKI, L. LE\'SNIAK}

\address{Department of Theoretical Physics, The Henryk Niewodnicza\'nski Institute of Nuclear Physics, Polish
Academy of Sciences, 31-342 Krak\'ow, Poland}

\author{A. P. SZCZEPANIAK}

\address{Physics Department and Nuclear Theory Center, Indiana University,
Bloomington, IN 47405, USA
}

\maketitle

 %\pub{Received (Day Month Year)}{Revised (Day Month Year)}

\begin{abstract}

We have analyzed the $\gamma p \to p K^+K^-$ reaction in the $K^+K^-$
effective mass region around the mass of the $\phi(1020)$ meson. The interference of the
$S$-wave contribution with the $P$-wave has been studied. Both scalar resonances
$f_0(980)$ and $a_0(980)$ have been taken into account. We obtained a good description
 of the available experimental data, in particular the mass distributions and the moments of
 the kaon angular distribution. Our calculations give values of the integrated $S$-wave 
 total photoproduction 
 cross section between 4 and
 7 nb for the $K^+K^-$ effective mass range around the  $\phi(1020)$ mass and at the laboratory
 photon energy near 5 GeV. These numbers
 favor lower experimental estimates obtained at DESY.
\end{abstract}
\section{Introduction}
The $K^+K^-$ photoproduction cross section is dominated by the production of the
$\phi(1020)$ meson. In
 the $S$-wave the near threshold $K^+K^-$ production is strongly influenced by
 the isoscalar $f_0(980)$ and
the isovector $a_0(980)$ resonances. The analysis of the near-threshold $K^+K^-$
dynamics is especially important for an explanation of the nature of the $f_0(980)$
meson. It is still a 
matter of
controversy whether this meson is genuine quasi bound $K\overline{K}$ state . 
Exploiting the large coupling of the photon to the vector mesons
along with the $S$-$P$ wave interference one is able to reveal many interesting
properties of the scalar mesons, for instance their photoproduction cross
section and their impact on the angular distribution of the produced kaons.
The data 
on the near threshold $K^+K^-$
photoproduction are, however, very scarce. Our analysis is based on the results obtained
by Behrend et al. at DESY \cite{Behrend} at the average incident photon energy of
5.65 GeV and by Barber et al. at Daresbury 
\cite{Barber} at 4 GeV. The main motivation of our study was to explain a huge discrepancy
in the $S$-wave cross sections reported by these experiments. The results varied between 2.7
and 96 nb respectively. We have also
performed a more extensive analysis of the moments of the angular distributions.
These moments are the main source of information about the partial wave interference.
To this end we have taken into account all the moments constructed from the $S$- and 
$P$-wave amplitudes. It is worth of mentioning that previous
analyses excluded all the moments but $\langle Y^1_0 \rangle$.
\section{Model description and numerical calculations}
Here we present the main ingredients of our model referring the reader to our
previous papers \cite{BLS,acta,Ji} for a more thorough study.
The partial wave projected amplitudes have been constructed for the reaction 
$\gamma p \to p K^+K^-$:
\begin{equation}
T_{\lambda_\gamma \lambda \lambda'}(t,\M,\Omega) = \sum_{L=S,P;M}   
T^L_{\lambda_\gamma \lambda \lambda' M}(t,\M)~ Y^L_{M}(\Omega),   
\end{equation}
where $\lambda_\gamma, \lambda, \lambda'$ are helicities of the photon, incoming and
outgoing protons, $\Omega$ is the $K^+$ solid angle , $t$ is 4-momentum transfer squared,
$\M$ is the $K^+K^-$ effective mass and
$L, M$ describe the angular momentum of the outgoing $K^+K^-$ system and its projection.
\vspace{-.5cm}
\begin{figure}[h]
%  \begin{minipage}[t]{.45\textwidth}
  \begin{center}
  \epsfig{file=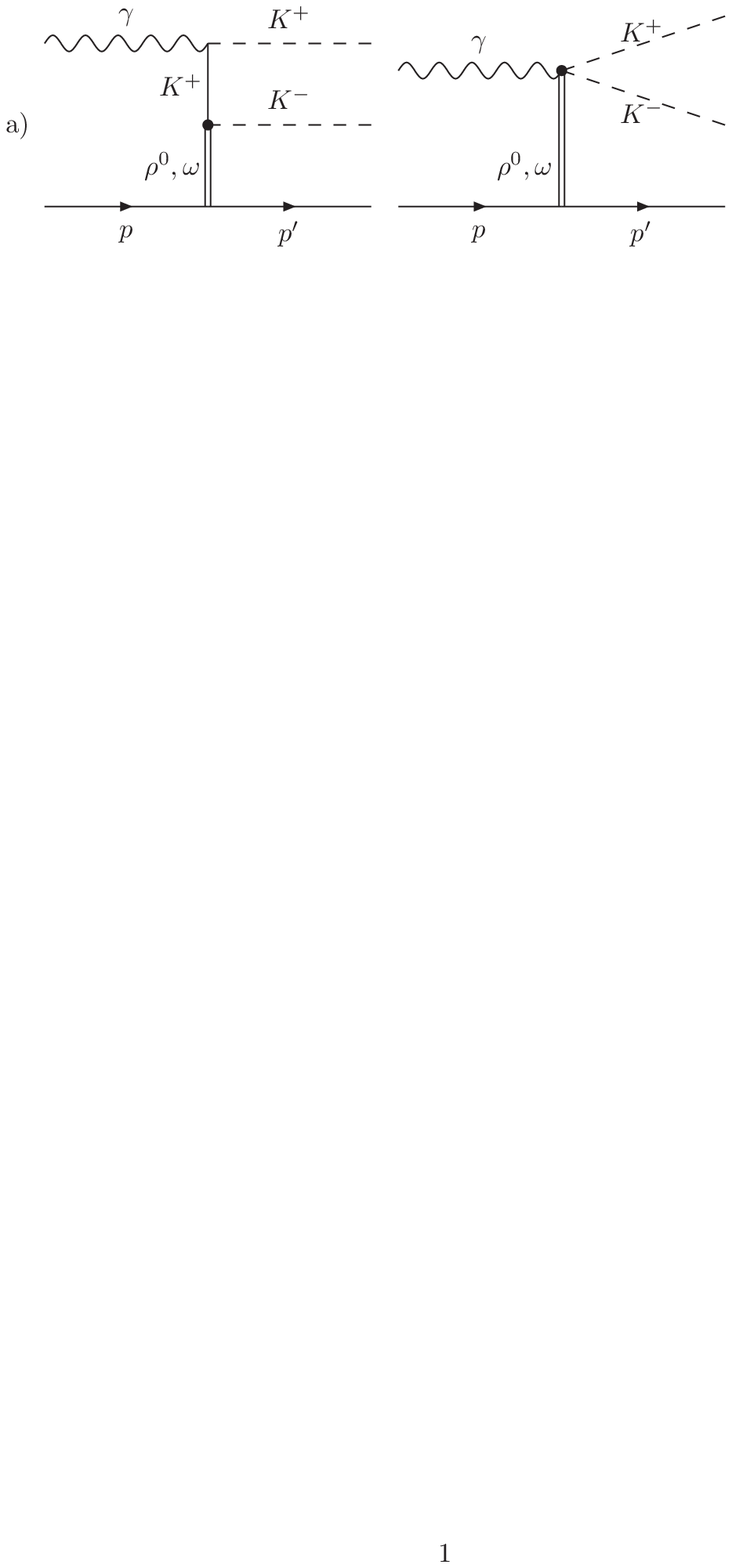, scale=.8}
%  \caption{Some diagrams representing the $K^+K^-$ Born amplitude.}
%  \end{center}
%  \end{minipage}
%  \hfill
%  \begin{minipage}[t]{.45\textwidth}
%  \begin{center}
  \epsfig{file=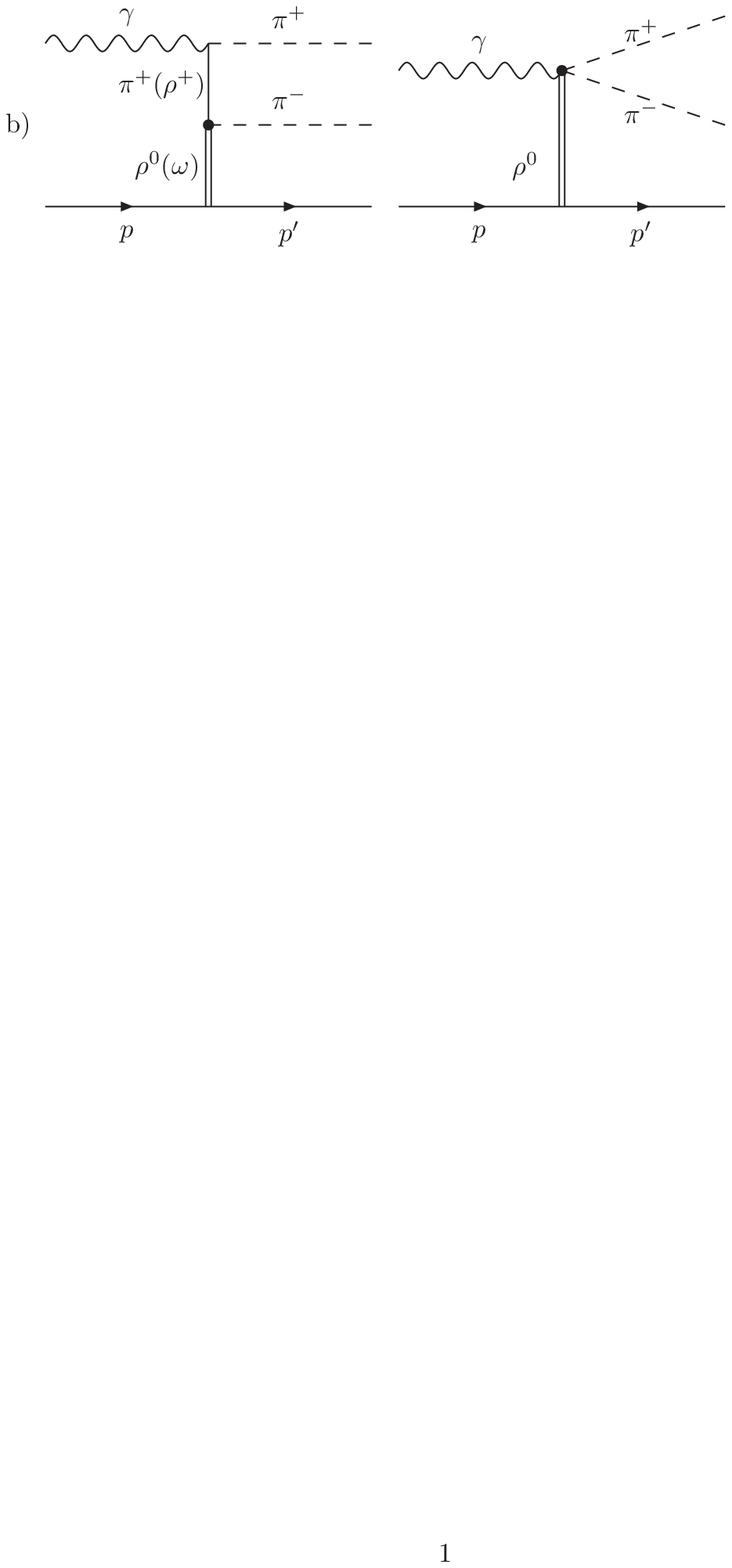, scale=.8}
  \caption{Some diagrams representing the $K^+K^-$ (a) and the $\pi^+\pi^-$ 
  (b) Born amplitudes}
  \end{center}
%  \end{minipage}
%  \hfill
\end{figure}
For the $S$-wave we have used the Born amplitudes schematically depicted in 
Fig. 1. Altogether we have  6 diagrams of the type presented in Fig.1a 
and 5 diagrams of the type presented in Fig. 1b.
The model has also taken into account the final state interactions. The
$\pi^+\pi^-\to K^+K^-$ and $\pi^0\pi^0\to K^+K^-$ transitions have been included along 
with the $K^+K^-$ elastic
rescattering. To describe the propagation of the exchanged particles we
have used either the normal $1/(t-m^2)$
or the Regge propagators
\begin{equation}
-[1-e^{-i\pi
\alpha(t)}]\Gamma(1-\alpha(t))(\alpha's)^{\alpha(t)}/(2s^{\alpha_0}),
\end{equation}
where $m$ is the mass of the exchanged particle and $\alpha(t)$ is its Regge 
trajectory.\\

In the $P$-wave we have assumed the reaction to proceed through the pomeron
exchange.
% as depicted at Fig. 3.
%\begin{figure}[h]
%\begin{center}
%\includegraphics[width=.6\textwidth]{graf_p.eps}
%\caption{The diagram describing the pomeron exchange in the $P-$ wave amplitude.}
%\end{center}
%\end{figure}
Having performed the numerical calculations we have arrived at a very good description of the mass
distributions and the moments which are shown in Fig. 2 for the case of
$E_\gamma$=4 GeV.
\vspace{.5cm}
\begin{figure}[h]
\begin{center}
\includegraphics[width=.7\textwidth]{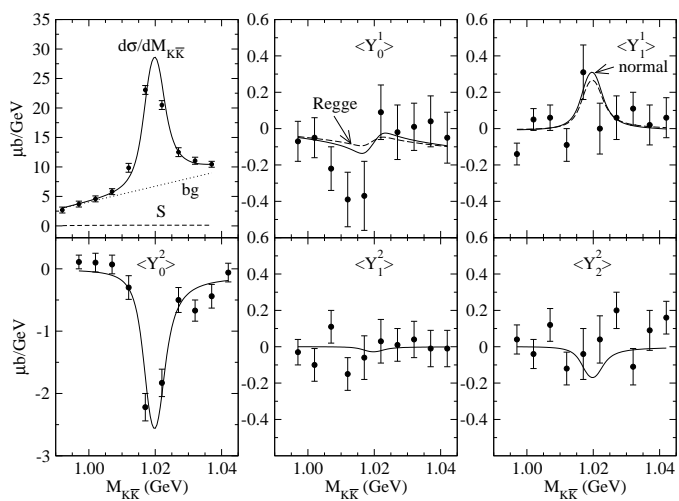}
\caption{The mass distribution and the moments for $E_\gamma$=4 GeV}
\end{center}
\end{figure}
Our model 
gives the value of the integrated \mbox{$S$-wave} photoproduction cross section in
the range between 4 and 7 nb for the photon energy varying from 4 GeV to 5.65 GeV .
\section{Summary}
Using the model constructed by us we have
 properly reproduced the mass distributions and the moments measured in the
experiments. Some other observables like the differential cross
section  $d\sigma/dt$ and the density matrix elements can also be described in agreement with
experimental data. Our calculation favors the DESY experimental estimation of the S-wave 
integrated total cross section.


\begin{thebibliography}{5}
\bibitem{Behrend} H.-J. Behrend {\it et al.}, Nucl. Phys. B{\bf 144},   
  22 (1978)    
   
\bibitem{Barber} D. P. Barber {\it et al.}, Z. Phys. C{\bf 12}, 1 (1982)   
 
\bibitem{BLS} \L.~Bibrzycki, L.~Le\'sniak and A.~P.~Szczepaniak,
Eur.\ Phys.\ J.\ C {\bf 34}, 335 (2004)

\bibitem{acta} L. Le\'sniak, A. P. Szczepaniak, Acta. Phys. Pol. {\bf B34},   
3389 (2003)  

\bibitem{Ji} C.- R. Ji, R. Kami\'nski, L. Le\'sniak, A. P. Szczepaniak,         
R. Williams, Phys. Rev. C{\bf 58}, 1205 (1998)


\end{thebibliography}
\end{document}